# Key signal contributions in photothermal deflection spectroscopy

Walter Dickmann[1,2], Johannes Dickmann[2], Florian Feilong Bruns[1] and Stefanie Kroker[1,2]

**AFFILIATIONS**

[1] Technische Universität Braunschweig, LENA Laboratory for Emerging Nanometrology, Am Langen Kamp 6, 38106 Braunschweig, Germany

[2] Physikalisch-Technische Bundesanstalt Braunschweig, Bundesallee 100, 38106 Braunschweig, Germany

*Preprint Manuscript*

**ABSTRACT**

We report on key signal contributions in photothermal deflection spectroscopy (PDS) of semiconductors at photon energies below the bandgap energy and show how to extract the actual absorption properties from the measurement data. To this end, we establish a rigorous computation scheme for the deflection signal including semi-analytic raytracing to analyze the underlying physical effects. The computation takes into account linear and nonlinear absorption processes affecting the refractive index and thus leading to a deflection of the probe beam. We find that beside the linear mirage effect, nonlinear absorption mechanisms make a substantial contribution to the signal for strongly focussed pump beams and sample materials with high two-photon absorption coefficients. For example, the measured quadratic absorption contribution exceeds *5*% at a pump beam intensity of about $1.3 \times 10^5$ W/cm$^2$ in Si and at $5 \times 10^4$ W/cm$^2$ in GaAs. In addition, our method also includes thermal expansion effects as well as spatial gradients of the attenuation properties. We demonstrate that these effects result in an additional deflection contribution which substantially depends on the distance of the photodetector from the readout point. This distance dependent contribution enhances the surface related PDS signal up to two orders of magnitude and may be misinterpreted as surface absorption if not corrected in the analysis of the measurement data. We verify these findings by PDS measurements on crystalline silicon at a wavelength of 1550 nm and provide guidelines how to extract the actual attenuation coefficient from the PDS signal.

## I. INTRODUCTION

Optical absorption processes in dielectric materials play an important role for various optical devices. These range, for example, from optical waveguides for telecommunication, solar cells for photovoltaics and nano-optical polarizers for the ultraviolet range up to components for high-precision interferometers and laser resonators.[1-6] Collinear photothermal deflection spectroscopy (PDS) is a powerful method to investigate spatially resolved absorption properties of transparent samples, e.g. organic materials as thin films and semiconductors below the band gap energy.[7-10] Particularly, in high-precision optical metrology, crystalline semiconductors such as silicon and gallium arsenide are promising materials. Thanks to their low mechanical losses these crystalline materials can provide a superior thermal noise performance of optical components.[11-15] Recent measurements of the attenuation coefficient of highly pure silicon samples indicate a significantly enhanced attenuation coefficient close to the sample surface, depending on the surface polishing procedure.[16,17] This enhanced surface absorption might be an important issue for silicon substrates as well as for structured high-reflectivity surfaces in which the surface area is even larger than in planar substrates.[18-21] The physical reasons for this enhancement are still being discussed.[17] Due to this apparently important role of the surface, it is necessary to pay attention to signal contributions in the PDS setup that predominantly appear close to the surface. Jackson *et al.* gave an analytical description of the setup, where, however, several important signal contributions were neglected, approximated or not discussed.[22]

In this work we present a method for the extraction of attenuation coefficients from PDS signals including all relevant signal contributions. To reliably quantify these contributions, we establish a rigorous computation scheme for the PDS signal that considers physical effects changing the refractive index itself and effects affecting the light propagation path. First, in section II we briefly introduce the principles of collinear photothermal deflection spectroscopy. In section III, we numerically analyze the PDS setup taking into account the mirage effect, free carrier creation and pure field effects. We then semi-analytically compute the probe beam deflection in the bulk material and close to the surface, additionally considering the thermal expansion of the sample and illustrate the implications of the computation results for the PDS experiment exemplary in a p-doped silicon sample at a wavelength of 1550 nm. In section IV we present the experimental results and illustrate how to extract the optical attenuation coefficient form the measured PDS signal.

## II. PRINCIPLES OF PHOTOTHERMAL DEFLECTION SPECTROSCOPY

FIG. 1 illustrates the measurement principle of photothermal deflection spectroscopy (PDS), a pump-probe setup for the spatially resolved determination of the optical attenuation coefficient.[7] A high power pump beam at the wavelength of interest is directed through the sample where it is partially absorbed. The pump beam is coupled into the sample at the Brewster angle $\theta_B$ to minimize the reflected pump power at the incoming and outcoming point and to maximize the total power deposition. The

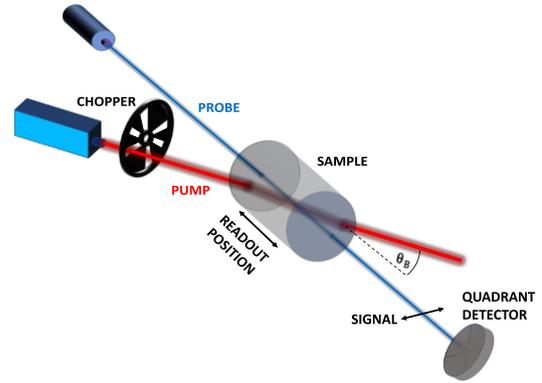

**FIG. 1.** Schematic illustration of a collinear PDS setup with a cylindrical sample. The local refractive index modulation induced by the absorption of the pump beam leads to a deflection of the probe beam that is measured by the quadrant detector.

Brewster angle is computed out of the refractive indices of the sample $n_z$ and air $n_0$:[23]

$$\theta_B = \arctan(n_z/n_0). \qquad (1)$$

The absorbed pump power leads to a local modulation of the refractive index due to mechanisms such as local heating and free carrier creation. The amplitude of this refractive index modulation is a measure for the local attenuation coefficient. In the PDS setup the index modulation is read out by a low power probe beam. The probe beam crosses the pump beam at the readout point that can be changed by a displacement of the sample. The absorption induced refractive index modulation leads to a deflection of the probe beam that is measured by a quadrant detector. An optical chopper harmonically modulates the pump beam power and therefore also the deflection signal. A corresponding lock-in amplifier enhances the signal-to-noise ratio.

## III. PDS SIGNAL COMPUTATION

The computation scheme of the PDS signal for semiconductors below the band gap energy is illustrated in FIG. 2. To compute the probe beam path (which is directly connected to the deflection signal), it is necessary to determine the refractive index field in the sample. In addition, thermal expansion can lead to changes of the sample shape. These sample shape changes as well as the refractive index changes are mainly caused by local heating due to the phonon-assisted absorbed pump power. In addition to that, other mechanisms like free carrier creation and pure field effects also change the refractive index field.[24] The influence of these additional effects on the refractive index modulation strongly depends on the material as well as experimental parameters such as pump beam intensity or sample geometry. In contrast to Jackson *et al.*[22], our computation scheme includes all discussed effects on the refractive index in semiconductor samples of virtually arbitrary geometries. In the following sections, we execute this computation exemplary for a cylindrical p-doped silicon sample.

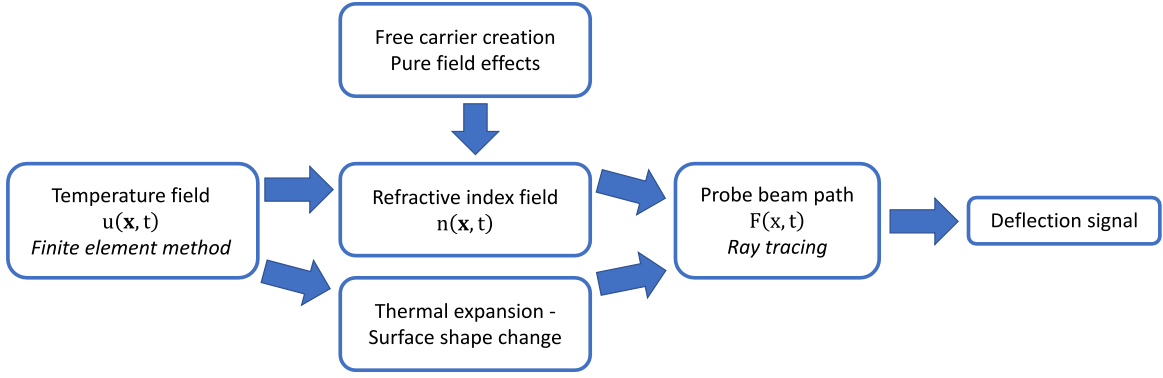

**FIG. 2.** Steps for the PDS signal computation.

## A. Temperature field in the PDS setup

As introduced in the previous section, the main contribution to the PDS signal is caused by the temperature induced refractive index change. To compute the temperature field in the sample, the heat equation[25]

$$\partial_t u(\boldsymbol{x},t) - a\,\Delta u(\boldsymbol{x},t) = f(\boldsymbol{x},t) \quad (2)$$

is solved for the experimental setup shown in FIG. 1. Here, $u$ is the temperature, $a$ is the thermal diffusivity and $f$ is the volumetric heat source. For a typical PDS configuration, the heat source is linked to the linear intraband absorption:[26]

$$f(\boldsymbol{x},t) = \frac{\alpha(x)I(x,t)}{\rho \cdot c_\mathrm{m}}. \quad (3)$$

The parameter $\alpha$ is the linear attenuation coefficient, $I$ is the pump beam intensity, $\rho$ is the volumetric mass density and $c_m$ is the mass heat capacity of the sample. We assume a Gaussian shaped pump beam which propagates along the $z$-direction and is focussed by a converging lens on a specific position into the sample (compare FIG. 3). The optical chopper periodically modulates the pump beam intensity $I(\boldsymbol{x},t)$. The PDS setup is established in a temperature stable environment, thus the initial condition is a constant temperature. The boundary condition on the sample surface takes into account conduction, convection and radiation. Further details are given in Appendix A. The heat equation (2) with initial and boundary conditions for the given sample geometry as shown in FIG.3 is solved by a finite element method (FEM). Therefore, a mesh is established which discretizes the sample region. More details of the FEM routine are given in Appendix A including a table that lists all input parameters which are used for the temperature field computation. FIG. 4 shows the spatial and temporal temperature field at the pump beam focus for a cylindrical p-doped crystalline silicon sample (conductivity of 3.0 Ωcm). All relevant experimental parameters (e.g. sample dimensions and pump intensity) are listed in TABLE I, Appendix A.

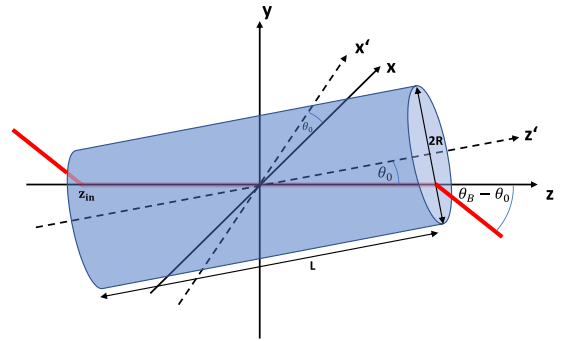

**FIG. 3.** Geometry of the cylindrical sample (radius R, length L). The pump beam hits the sample at the Brewster angle $\theta_B$ relative to the cylinder axis z'. The angle between the refracted pump beam in the sample which defines the z-axis and the cylinder axis is $\theta_0 = \pi/2 - \theta_B$.

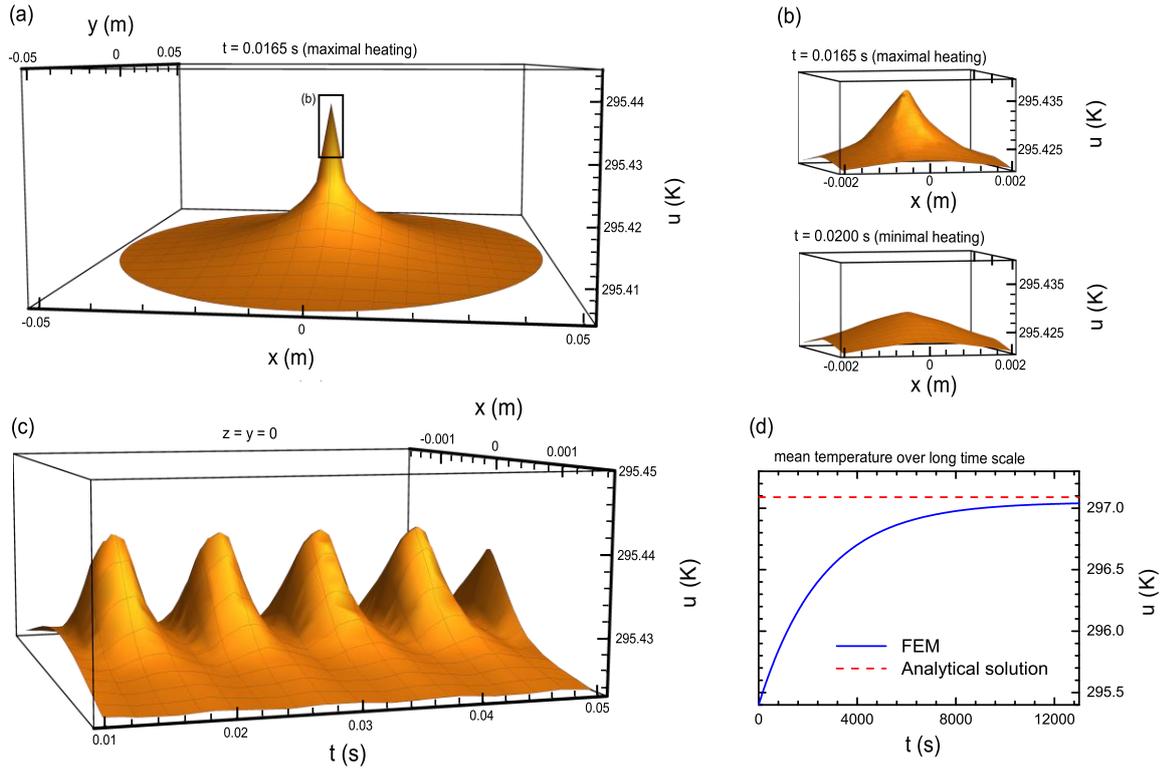

**FIG. 4.** Computed spatial and temporal temperature field at the pump beam focus for a cylindrical p-doped crystalline silicon sample (computation parameters in TABLE I, Appendix A). (a) Spatially resolved temperature field for $t = 0.0165$ s (maximal heating) and $z = 0$. (b) Temperature field close to the beam centre for $t = 0.0165$ s (upper figure, maximal heating) and $t = 0.0200$ s (lower figure, minimal heating). The maximum temperature difference is 11.5 mK. (c) Time and $x$-resolved temperature field ($z = y = 0$). (d) Mean sample temperature over a long time scale (blue line). The red line is the analytically calculated equilibrium temperature for $t \to \infty$.

The maximum temperature difference during one chopper period ($f_{Ch} = 108$ Hz) is approximately 11.5 mK in the pump beam focus. Over a typical measurement time of about 300 s, the mean sample temperature is increased by approximately 150 mK (see FIG. 4 d). For a typical wafer, this value would be even smaller as it decreases with decreasing sample length due to the reduced heat deposition. This long time temperature enhancement has no significant influence on the following PDS signal computation. However, for samples with smaller diameters or higher absorption coefficients, the sample heating might be relevant in terms of temperature dependent optical constants. Silicon has a high thermal diffusivity which leads to a fast broadening of the heating profile. Thus, the area of temperature modulation is with approximately $(1.5 \times 1.5)$ mm$^2$ about two orders of magnitude larger than the pump beam intensity area with $(0.1 \times 0.1)$ mm$^2$. In the next section we will illustrate how this temperature profile influences the refractive index distribution in the sample.

### B. Refractive index field

The determination of the deflection of the probe beam requires the knowledge of the refractive index field $n(\boldsymbol{x}, t)$ in the sample. In the following we discuss the main contributions to the refractive index change for semiconductors and photon energies below the band gap. These contributions are heating effects due to linear intraband absorption $\Delta n_\text{u}$[27], free carrier creation by two-photon absorption (TPA) $\Delta n_\text{C}$[28] and pure field effects, e.g. the Kerr effect $\Delta n_\text{E}$[29] resulting in:

$$n(\mathbf{x},t) = n_\text{z} + \Delta n_\text{u}(\lambda, u(\mathbf{x},t)) + \Delta n_\text{C}(\lambda, I(\mathbf{x},t)) + \Delta n_\text{E}(I(\mathbf{x},t)). \quad (4)$$

Linear intraband absorption mainly leads to heating of the sample which was computed in the previous section. The refractive index $n_\text{u}(\lambda, u)$ is computed by a Sellmeier model.[27] In the relevant temperature range (modulation amplitude about 15 mK, see FIG. 4) a linear approximation of the refractive index change at the probe beam wavelength $\lambda$ is possible:

$$\Delta n_\text{u}(\lambda, u) = \frac{dn}{du}(\lambda, u)\, \Delta u. \quad (5)$$

Besides that, interband absorption predominantly leads to a refractive index change due to the creation of free electrons by TPA. Using the electron lifetime $\tau_\text{C}$ in the sample and taking into account $f_\text{Ch}^{-1} \gg \tau_\text{C}$, the change of carrier density $\Delta N_\text{C}$ due to TPA is computed as[28]

$$\Delta N_\text{C} = \frac{\tau_\text{C} \lambda_\text{P} \beta I^2}{2 h c_0}. \quad (6)$$

Here, $h$ is the Planck constant, $c_0$ is the vacuum speed of light and $\beta$ is the two-photon absorption coefficient. The resulting refractive index change $\Delta n_\text{C}$ is estimated by a Drude model:[30]

$$\Delta n_\text{C}(\lambda, I) = -\frac{e^2 \lambda^2}{8\pi^2 c_0^2 \varepsilon_0 n_\text{z}} \left( \frac{\Delta N_\text{C}}{m_\text{e}^*} + \frac{\Delta N_\text{C}}{m_\text{h}^*} \right). \quad (7)$$

The probe beam wavelength is $\lambda$, $e$ is the elementary charge, $\varepsilon_0$ is the vacuum permittivity, $m_\text{e}^*$ is the effective mass of the electrons and $m_\text{h}^*$ is the effective mass of the holes in the sample. Note that this carrier induced refractive index change and thus the corresponding absorption signal is an upper estimate, as the diffusion length of the free carriers might be larger than the intensity profile of the pump beam for relatively pure samples.[31] Thus, diffusion processes might lead to a broader and lower amplitude refractive index profile.

Different pure field effects may influence the refractive index of semiconductors, e.g the Pockels effect, the Kerr effect and the Franz-Keldysh effect.[24,32-34] Depending on the material and the wavelength and intensity regime, it is necessary to account for the most relevant ones. For strain-free crystalline silicon with the experimental parameters shown in TABLE 1, Appendix A, the Kerr effect is the most relevant pure field effect. Its influence on the refractive index is estimated by the anharmonic oscillator model by Moss *et al.*:[29]

$$\Delta n_\text{E}(\lambda) = -\frac{3e^2(n_\text{z}^2 - 1)I}{c_0 \varepsilon_0 n_\text{z}^2 m_\text{e}^2 \omega_0^4 X^2}. \quad (8)$$

Here, $m_\text{e}$ is the electron mass, $\omega_0$ the oscillator resonance frequency and $X$ the average oscillator displacement. Note that, for the sake of simplicity, this computation takes only the main mechanisms into account that contribute to refractive index changes for the present experimental parameters. Other effects, like Franz-Keldysh effect or a change of $\alpha$ due to $\Delta N_\text{C}$, are neglected due to their small influence on the refractive index. Nevertheless, these effects might be relevant for silicon at other wavelengths, e.g. near bandgap, or intensity regimes and they are well documented in literature, e.g. by Soref and Bennet.[24] Also the self-focusing of the pump beam due to the thermal lens effect is neglectable for the present experimental parameters. However, it is relevant for higher pump powers or absorption coefficients. The self-focusing effect is well documented by Dabby and Whinnery.[35] The input parameters for the computation of the refractive index field are listed in TABLE II, Appendix A. Whereas the spatial profiles $\Delta n_\text{C}(x)$ and $\Delta n_\text{E}(x)$ are analytically given by equations (7) and (8), the profile $\Delta n_\text{u}(x)$ (see equation (5)) is based on the numerically computed temperature field. FIG 5 (a) shows the results for $\Delta n_\text{u}(x)$ and $\Delta n_\text{C}(x)$ for the silicon sample. The Kerr effect $\Delta n_\text{E}$ is about 5 orders of magnitude smaller and therefore not illustrated.

### C. Bulk deflection

Using the refractive index field computed in section III B, the probe beam path through the sample is computed by ray tracing.[36] To get an insight into relevant refraction processes that may contribute to the deflection signal, we discuss the most important details here. The refraction of the probe beam is described by Snell's law:

$$n(x,t)\sin(\varphi(x,t)) = n(x_{in},t)\sin(\varphi_{in}(t)) \equiv NA(t). \quad (9)$$

Here, $n(x_{in}, t)$ is the refractive index of the substrate at the entrance point $x_{in}$ of the probe beam and $\varphi(x,t)$ is the local refraction angle. The refraction angle at the entrance point $\varphi_{in}(t)$ defines the numerical aperture $NA(t)$. The resulting probe beam path reads as:

$$F(x,t) = \int \left[\sqrt{\frac{NA(t)^2}{n(x,t)^2-NA(t)^2}} + \frac{\partial_z n(x,t)}{\partial_x n(x,t)}\right] \times \left[1 - \sqrt{\frac{NA(t)^2}{n(x,t)^2-NA(t)^2}} \cdot \frac{\partial_z n(x,t)}{\partial_x n(x,t)}\right]^{-1} dx. \quad (10)$$

The numerical aperture NA is defined by the angle of the probe beam relative to the z'-axis. In the present case the Brewster angle is used (see FIG. 14, Appendix B). A smaller crossing angle between the beams would increase the PDS signal[22], but also decrease the spatial resolution. Further details of the bulk beam path computation are given in Appendix B. To illustrate the contributions of the different mechanisms changing the refractive index (compare previous section) on the deflection, the probe beam paths at the time of maximum refractive index change (i.e. during maximum deflection) are computed by considering either only $n_u$, $n_C$ or $n_E$. The computation of $F(x,t)$ with equation 10 gives the probe beam paths illustrated in FIG. 5 (b). Here, the parallel bulk deflection $d_{bulk,0}$ is plotted symmetrically over the probe axis position. The resulting deflection induced by the temperature field is with approximately 3.9 nm about 13 times larger than the free carrier induced deflection of 0.3 nm (see FIG. 5 (b)). Therefore, we expect a small but measurable nonlinear contribution to the PDS signal (see section IV C). In total, a parallel deflection $d_{bulk,0}$ of about 4.2 nm occurs due to refractive index gradients in the bulk for the present experimental parameters.

It is important to note that the computation above is based on the approximation of a perfect symmetrical refractive index field with respect to the pump beam axis z. Indeed, the temperature field (and thus the refractive index field) is not exactly symmetric to the z-axis. One reason for that is the chopping of the pump beam which destroys its symmetrical gaussian shape (see Appendix A). Another possible reason is a spatial variation of the optical attenuation $\alpha$ along the probe beam axis. The resulting asymmetry in $\Delta n_u(x)$ causes an additional angular change $\Delta\varphi_{bulk,1}$ of the probe beam in the sample. Using Snell's law, the resulting angular change $\Delta\varphi_{bulk,2}$ of the beam behind the sample is given by:

$$\Delta\varphi_{bulk,2} = \frac{\sqrt{n_z^2 - n_0^2 \sin^2\varphi'_{in}}}{n_0 \cos\varphi'_{in}} \Delta\varphi_{bulk,1}. \quad (11a)$$

The total bulk deflection $d_{bulk}$ at a given detector distance $D_{det}$ for read out at a distance $D_m$ from the front surface consists of angular and parallel deflection contributions:

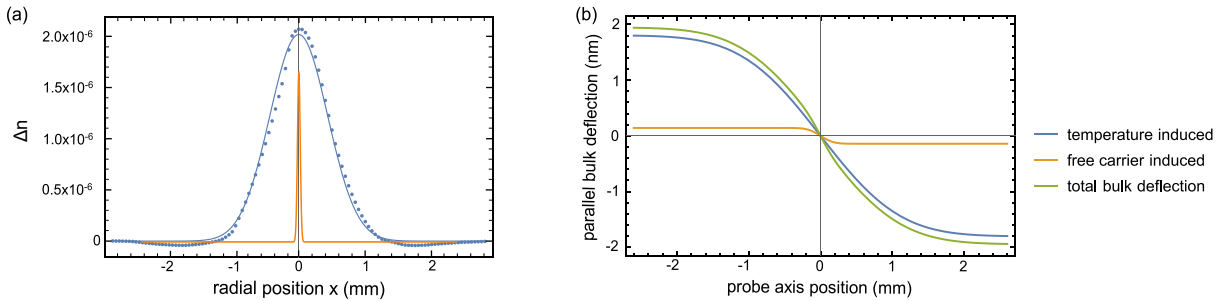

**FIG. 5.** (a) Refractive index profiles $\Delta n_u(x)$ and $\Delta n_C(x)$ over the radial position $x$ in the sample. A Gaussian fit $\Delta n_u(x) = \Delta\hat{n} \times \exp[-x^2/\sigma_n^2]$ of the temperature induced refractive index profile in the sample gives the fit parameter values $\Delta\hat{n} = 2.02 \times 10^{-6}$ and $\sigma_n^{-2} = 2.64 \times 10^6$ m$^{-2}$. (b) Probe beam paths through the symmetrically approximated index gradient field. The probe axis position denotes the distance from the crossing point between the two beams along the probe beam propagation direction.

$$d_{\text{bulk}} = d_{\text{bulk},0} + \frac{L-D_{\text{m}}}{\cos\theta_0}|\Delta\varphi_{\text{bulk},1}| + D_{\text{det}}|\Delta\varphi_{\text{bulk},2}|. \quad (11b)$$

According to equation (11b) the angular and parallel deflection contributions can be separated experimentally by distance dependent measurements. In this way, in section IV we will demonstrate that the angular deflection effects can contribute significantly to the overall deflection signal and consequently should be taken into account in absorption measurements with PDS.

**D. Surface deflection**

Beside the refractive index modulation mechanisms and the angular effects described in section III C, two additional effects have a main impact onto the deflection signal at the sample surface: First, the oscillating temperature field around the pump beam leads to a periodic thermal expansion and contraction of the sample surface. The total expansion along the pump beam axis (in $z$-direction) reads as

$$\Delta z(x,y,t) = \int_{-z_{\text{in}}}^{z_{\text{in}}} \alpha_{\text{L}}\big(u(\boldsymbol{x},0) - u(\boldsymbol{x},t)\big)\,\mathrm{d}z. \quad (12)$$

For crystalline silicon, the linear expansion coefficient is $\alpha_{\text{L}} = 2.5 \times 10^{-6}$ K$^{-1}$.[37] Neglecting radial expansion and stresses, the maximum surface expansion given in FIG. 6 as a function of the radial position $x$ on the surface follows (computation details are shown in Appendix C). A maximum expansion amplitude of about 1.6 nm is expected. This thermal expansion leads to an angular deflection contribution that is computed in the following section. The thermal expansion changes the refraction angle of the probe beam and thus causes an additional distance dependent deflection.

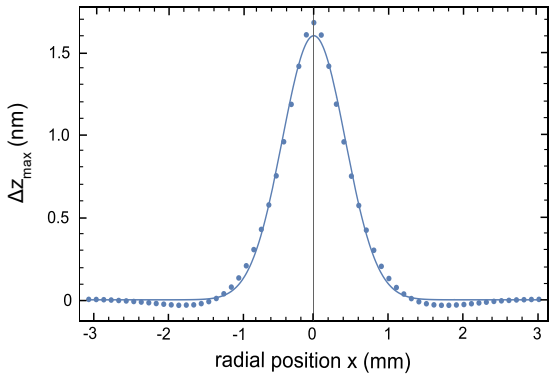

**FIG. 6.** Computed temperature induced back surface expansion as a function of the distance $x$ from the pump beam axis. The surface shape change is described by a gaussian function $\Delta z_{\max}(x) = \Delta \hat{l} \times \exp[-x^2/\sigma_z^2]$ with $\Delta \hat{l} = 1.60 \times 10^{-9}$ m and $\sigma_z^{-2} = 2.87 \times 10^6$ m$^{-2}$ as fit parameters. At the front surface a similar thermal expansion effect occurs.

As a second effect, the probe beam path (see FIG. 5(b)) shows a significant bending close to the pump beam axis. Probing close to the surface, this bending also modulates the refraction angle and leads to an additional distance dependent deflection. The total probe beam deflection is again computed by Snell's law. Using small angle approximations, the total front surface deflection reads as:

$$d_{\text{FS}} = \exp\left[\alpha\frac{L}{2\cos\theta_0}\right]d_{\text{FS},0} + \frac{L}{\cos\theta_0}|\Delta\varphi_{\text{FS},1}| + D_{\text{det}}|\Delta\varphi_{\text{FS},2}|, \quad (13a)$$

with

$$\Delta\varphi_{\text{FS},1} = \frac{\mathrm{d}}{\mathrm{d}x}\Delta z_{\max}\left(\frac{n_0}{n_z}\tan\theta_0 - 1\right) + \exp\left[\alpha\frac{L}{2\cos\theta_0}\right]\frac{\mathrm{d}}{\mathrm{d}x}d_{\text{bulk},0}\sin(2\theta_0) \quad (13b)$$

and

$$\Delta\varphi_{\text{FS},2} = \frac{n_z}{n_0}\cot\theta_0 \Delta\varphi_{\text{FS},1}. \quad (13c)$$

Equation (13a) contains the probe beam deflection as discussed for the bulk as well as the deflection due to the additional surface effects that change the refraction angle. For the front surface, beside the parallel deflection contribution (first term in equation (13a)), there are two different modulation angles: $\Delta\varphi_{\text{FS},1}$ as the angle change of the probe beam during its way through the sample and $\Delta\varphi_{\text{FS},2}$ as the angle change behind the back surface. Again, these two angles are connected by Snell's law.

The total back surface deflection reads as

$$d_{\text{BS}} = \exp\left[-\alpha\frac{L}{2\cos\theta_0}\right]d_{\text{BS},0} + D_{\text{det}}|\Delta\varphi_{\text{BS}}| \quad (14a)$$

with the angular modulation

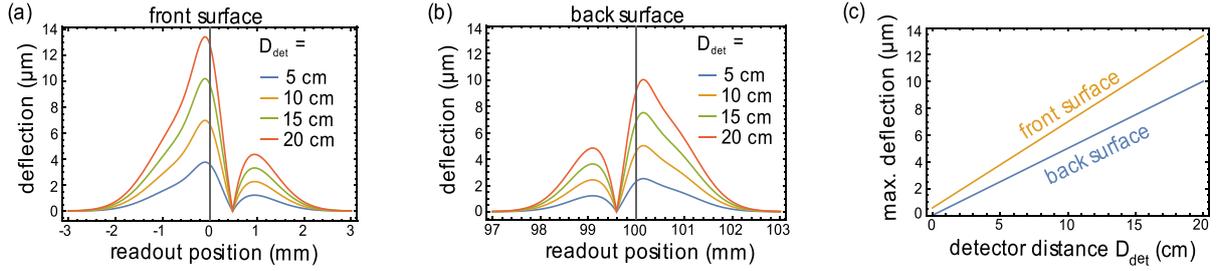

**FIG. 7.** Computed (a) front and (b) back surface deflection as a function of the distance $D_m$ between the readout point and the front surface for different detector distances $D_{det}$. The distance dependence of the maximum deflection is plotted in (c). For the computations we assumed the surface absorption to be equal to the bulk absorption.

$$\Delta\varphi_{BS} = \frac{n_z}{n_0}\cot\theta_0 \times \left(\frac{d}{dx}\Delta z_{max} + \exp\left[-\alpha\frac{L}{2\cos\theta_0}\right]\frac{d}{dx}d_{bulk,0}\sin(2\theta_0)\right). \quad (14b)$$

For the back surface deflection, there is only one angle modulation $|\Delta\varphi_{BS}|$. For both the front and the back surface, the angular and parallel deflection contributions can be separated experimentally just as for the bulk deflection. Further details of this computation are shown in Appendix C.

The total deflections $d_{FS}$ and $d_{BS}$ are plotted in FIG. 7 (a) and (b) for different detector distances over the readout position $D_m$, which has already been introduced in section III C. The dependence on the readout position is encoded in equations (13a)-(14b) in the spatial variable x (see equation (C9) in Appendix C). Both figures show that angular effects lead to a strongly enhanced deflection probing close to the sample surface. The maximum deflections for the front and back surface are plotted in FIG. 7 (c) over the detector distance $D_{det}$. In agreement to equations (13a) and (14a) it illustrates the linear increase of the PDS surface signal with the detector distance. Importantly, the surface contributions can lead to a computed deflection enhancement of about 3 orders of magnitude which however is not directly related to the absorption in the material. For the computation, we assumed an infinitely small probe beam. A real probe beam, however, has an inherent divergence as it is focussed on the readout point. Furthermore, the refraction process is not uniform along the probe beam profile because of a spatial varying refractive index gradient. That causes an additional divergence of the probe beam during the readout process. Thus, increasing the detector distance leads to a reduced detector sensitivity and the expected signal enhancement due to the angular contributions will be smaller than the computed deflection enhancement. One can estimate this effect by considering the quadrant diode sensitivity (that is described e.g. by Manojlović[38]) and the probe beam divergence given in the experiment. In our case, this divergence leads to an up to 90 times reduced detection sensitivity for large detector distances compared to an infinitely small probe beam. Nevertheless, this approximation has no practical relevance for the analysis of actual PDS measurements, as the angular and parallel deflection can be separated experimentally by distance dependent measurements. This is shown in section IV D.

## IV. EXPERIMENTAL METHODS AND RESULTS

### A. Experimental setup

FIG. 8 shows a sketch of the collinear PDS setup for spatially resolved absorption measurements. The pump beam is provided by a cw fiber laser ($P_P = 5$ W, $\lambda_P = 1550$ nm) and the probe source is a laser diode ($P_1 = 1$ mW, $\lambda_1 = 1310$ nm). The reflected pump power is adjusted to zero (Brewster angle refraction) to minimize internal reflections and to maximize the absorbed pump power. To change the readout position a translation stage is used. To quantify the previously described angular signal contributions, the detector distance $D_{det}$ is varied. The calibration of the system is realized by an absolute transmission measurement which gives the total transmittance $T$. Surface shape changes due to thermal expansion are investigated by laser interferometry.

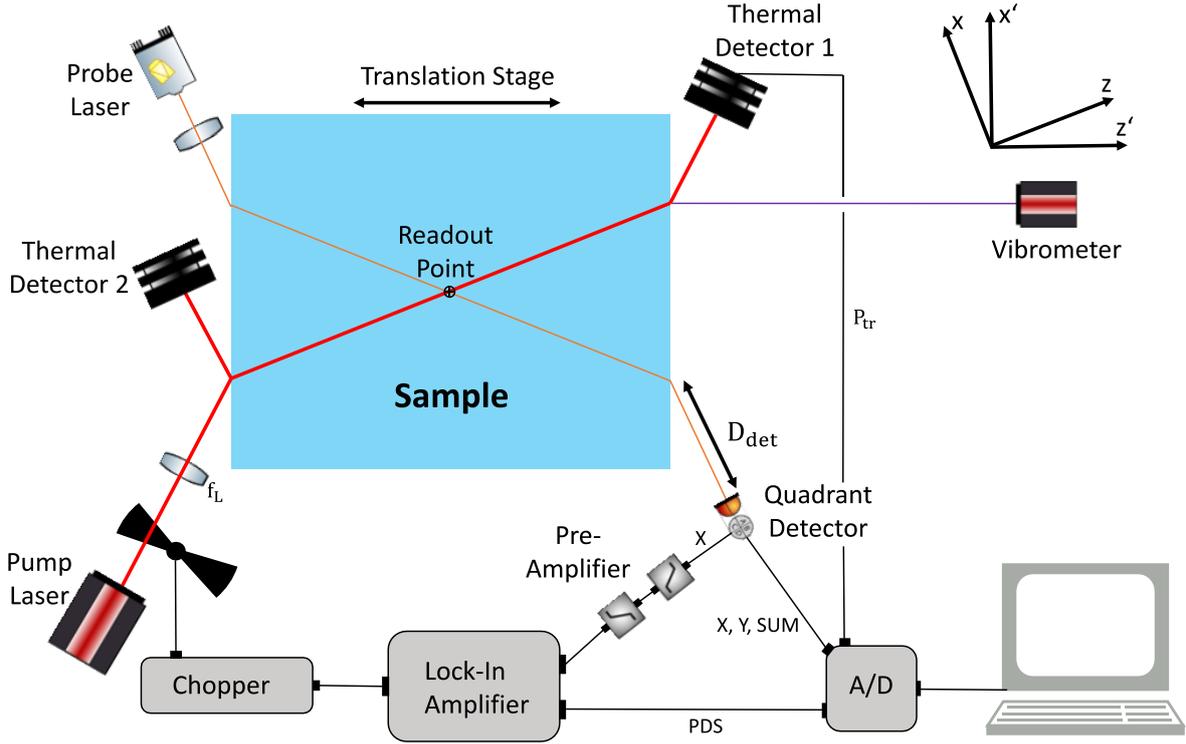

FIG. 8. Measurement setup for PDS and surface expansion readout. The pump beam is provided by a cw laser ($P_P = 5$ W, $\lambda_P = 1550$ nm) and intensity modulated by the optical chopper. Both reflected and transmitted pump power are measured with thermal detectors. The probe source is a diode laser ($P_1 = 1$ mW, $\lambda_1 = 1310$ nm). The probe beam deflection is measured with a quadrant detector. A direct readout of the surface expansion is realized with a laser vibrometer.

## B. Investigation of thermal expansion

Analogously to the theoretical analysis of PDS, we investigate the effects of thermal expansion exemplary in a p-doped c-silicon sample. The properties of the Si sample as well as the other relevant experimental parameters are given in TABLE I, Appendix A. The surface shape modulation is read out by a laser vibrometer (type *SIOS SB01*). The vibrometer beam is focused and guided perpendicular onto the back surface of the sample (see FIG 8). By moving the translation stage, the outcoupling point of the pump beam (and thus the expansion region) is shifted in *x'*-direction (parallel to the cylinder bases), while the reflection point of vibrometer beam has a constant position. This enables a scanning of the surface and measuring its expansion. FIG. 9 (a) shows the spectral vibrometer signal for three different readout positions. The frequency of maximum amplitude corresponds to the chopper frequency at which the pump laser light is modulated. The measured signal amplitude at the chopper frequency as a function of the surface position as well as the simulated surface expansion are shown in FIG 9 (b). Obviously, the computed surface expansion fits well to the measurement. The maximum expansion is about 2 nm. This leads to a change of the refractive angle and an additional angular deflection of the probe beam which will be discussed in section IV D.

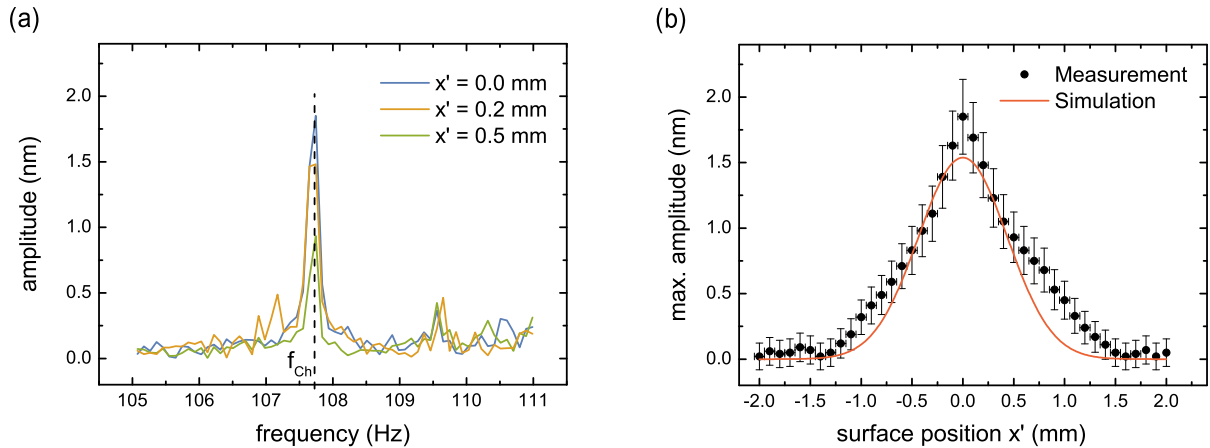

FIG. 9. $x'$- scan of the back surface expansion at y=0. (a) Spectral vibrometer signals at different surface positions $x'$. (b) Maximum signal amplitude as a function of $x'$ and computed surface expansion.

## C. Contribution of linear and quadratic absorption effects to the deflection signal

FIG. 10 shows the deflection signal in the bulk in silicon and gallium arsenide for different pump beam powers. The measurement data are fitted with a parabolic function to separate the contribution of linear and quadratic effects on the signal. In addition, FIGs. 10 (a) and (b) also show the influence of two different beam waists on the PDS signal. For silicon, linear absorption mechanisms are by far the most relevant contributions to the deflection signal in the present intensity regime. However, the quadratic absorption is measurable for high power and strongly focussed pump beams (i.e. high pump intensities). The inset in FIG. 10 (a) shows the quadratic contribution to the total PDS signal in silicon. It exceeds *5*% at a pump beam intensity of about $1.3 \times 10^5 \, \text{W/cm}^2$, respectively. The quadratic absorption contribution is much more pronounced in GaAs which is shown in FIG 10 (b). It exceeds 5% already at a pump beam intensity of about $5 \times 10^4 \, \text{W/cm}^2$, respectively. This is attributed to the about 13 times higher two-photon absorption coefficient in GaAs as compared with Si.[39] For comparison, the power dependence of the deflection signal has also been computed considering the effects described in section III. The contribution of nonlinear absorption mechanisms can be tested by PDS measurements at different pump beam powers. They are particularly relevant for strongly focussed pump beams and high pump powers. The handling of quadratic signal contributions is demonstrated in Appendix D.

## D. Extraction of the attenuation coefficient

To extract the attenuation coefficient, the angular deflection contributions must be separated from the PDS signal. FIG. 11 (a) shows the spatially resolved PDS signal for the silicon sample for three different detector distances. The insets show the signal around the sample centre, at the back surface and the air signal (outside the sample). First, we give an interpretation of FIG. 11 regarding the main features of the signals. Thereby we refer to the signal computation results given in section III D. The shapes of the measured surface signals qualitatively fit with the computed ones that are plotted in FIGs. 7 (a) and (b). The shoulders next to the main maxima at the front and back surface are a result of the thermal expansion. Without this effect, the surface signal shape would be approximately symmetrical. The widths of the measured surface signals are about 2 times larger than the computed ones, which may be caused by the finite size of the probe beam that is not considered in the computations. The surface related signal peaks exhibit a pronounced dependence on the detector distance $D_{det}$ demonstrating that the detector distance is a critical parameter that can substantially increase the PDS signal.

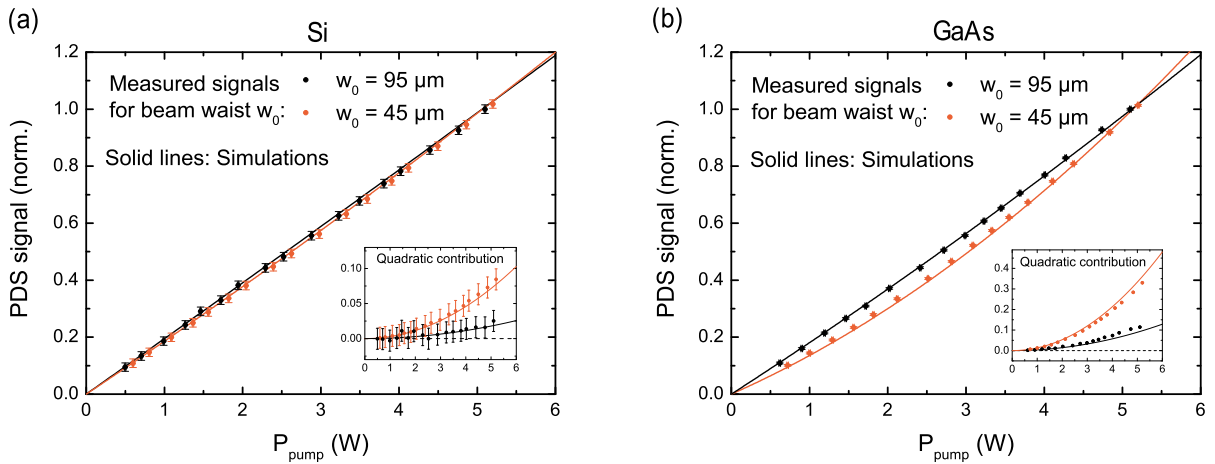

**FIG. 10.** Power dependence of the PDS signal in the bulk for (a) the silicon and (b) the GaAs sample for two different beam waists. The insets show the extracted quadratic contribution to the total PDS signal. The solid lines are simulation results and the dots are measurements.

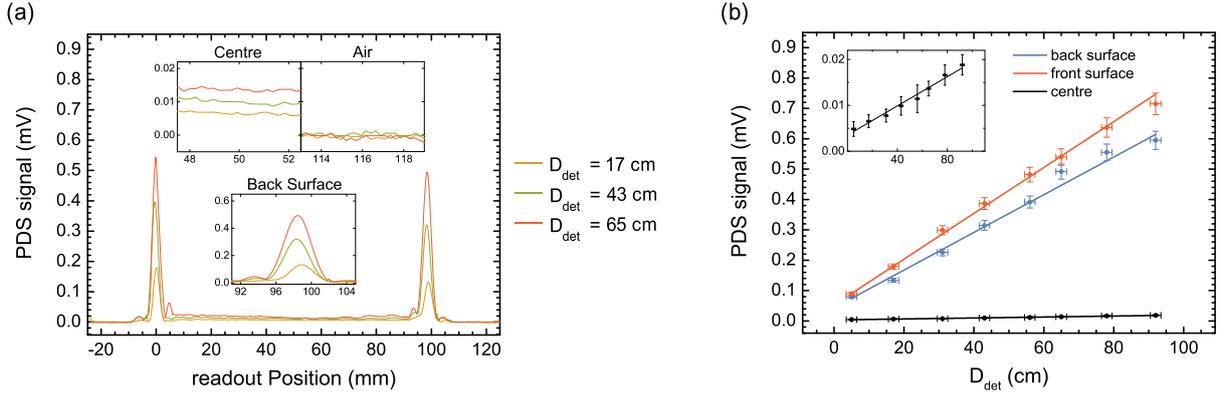

**FIG. 11.** (a) PDS signal over readout position along the silicon sample. The insets show the deflection signal in the centre of the sample, at the back surface and behind the back surface (air signal). (b) PDS signal over the detector distance for front and back surface and the sample centre. The inset shows the centre signal course with adapted scale.

The distance dependencies of the mean bulk signal and the maximum surface signals are shown in FIG. 11 (b), considering all measured spectra of the silicon sample. The slope of the front surface signal is with $(0.0075 \pm 0.0001)$ mV cm$^{-1}$ significantly higher than the value of the back surface of $(0.0062 \pm 0.0003)$ mV cm$^{-1}$. This agrees with the computed result plotted in FIG. 7 (c) and is caused by the attenuation of the pump beam on its way through the sample. The distance dependencies of both surface signals have a much higher slope than the bulk signal with $(0.00016 \pm 0.00001)$ mV cm$^{-1}$ which is attributed to the angular deflection effects that only occur close to the surface. However, the measured surface signal enhancements are smaller than the computed surface enhancements. That deviation is caused by the divergence of the probe beam after the readout process as described in section III D. The deviation has no influence on the attenuation extraction procedure. For the extraction of absorption data, it is essential that angular effects vanish at the respective readout point.

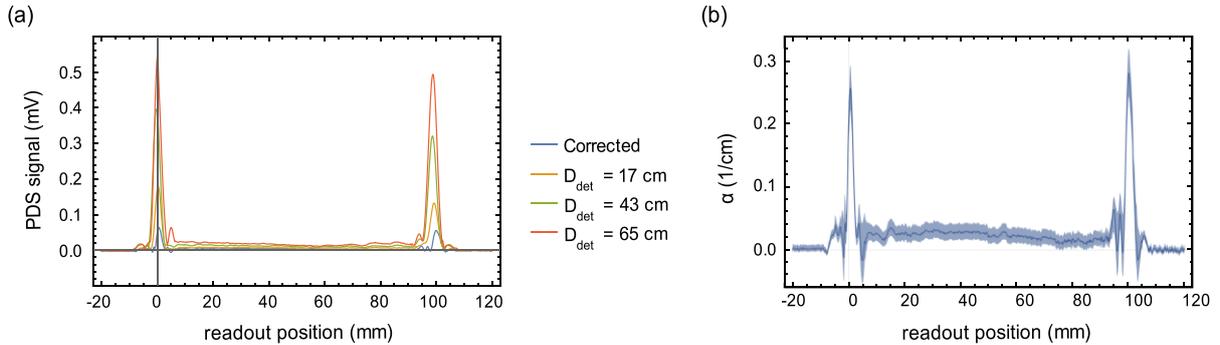

**FIG. 12.** (a) Extraction of the corrected PDS signal (silicon sample) by pointwise extrapolation to the readout point. (b) Attenuation coefficient extracted from the corrected signal.

From the distance dependent measurements, the attenuation coefficient can be extracted from the PDS signal. To this end, we determine a corrected deflection $g(z)$ by pointwise linear extrapolation of the distance dependent measurements back to the respective readout position, using equations (11), (13) and (14). For the silicon sample, the non-linear absorption mechanisms can be neglected in the measurement regime discussed above (see FIG. 10 (a)). Thus, the relation $g(z) \propto \alpha(z)P(z)$ between the PDS signal and the pump power is fulfilled. Hence, the linear attenuation coefficient is given by

$$\alpha(z) = g(z)\left[\frac{1}{1-T}\int_{\mathbb{R}} g(w)\mathrm{d}w - \int_{-\infty}^{z} g(w)\mathrm{d}w\right]^{-1}, \quad (15)$$

considering the total transmittance $T$ to calculate the integration constant. The measured transmittance is $T = 0.743 \pm 0.032$. FIG. 12 (a) shows the corrected deflection $g(z)$ and FIG. 12 (b) shows the resulting attenuation coefficient $\alpha(z)$.

The total error of $\alpha(z)$ is determined by the error of the corrected signal $g(z)$ (linear regression error) and by the total transmission error. The attenuation coefficient in the bulk material is in the range $[0.00 \dots 0.05]$ cm$^{-1}$ with a mean value of about $0.02$ cm$^{-1}$ which is in good agreement to literature values for the present doping concentration.[24] The maximum front surface value is $(0.258 \pm 0.034)$ cm$^{-1}$ and the maximum back surface value is $(0.278 \pm 0.037)$ cm$^{-1}$. Thus, the corrected signal leads to an up to 10 times enlarged surface absorption. This is much smaller than the uncorrected ratio between bulk and surface signal for large detector distances (e.g. factor 38 for $D_{\mathrm{det}} = 92$ cm) and shows that it is necessary to account for angular deflection contributions to obtain proper attenuation coefficients.

Note that for higher pump beam intensities and semiconductors with high two-photon absorption coefficients (compare section IV C), quadratic signal contributions may be relevant. Then, the relation $g(z) \propto \alpha(z)P(z) + \beta(z)P^2(z)$ results in coupled differential equations for different pump powers relating $\alpha(z)$, $\beta(z)$ and $g(z)$. In this case $\alpha(z)$ and $\beta(z)$ must be computed numerically. To this end, it is necessary to measure at least two different PDS signals $g(z)$ for two different pump beam powers to extract the linear and quadratic attenuation coefficients. Further details of the extraction formalism with quadratic absorption contributions are given in Appendix D.

**V. CONCLUSION**

In this contribution we report on key signal contributions and their experimental treatment in photothermal deflection spectroscopy (PDS) in semiconductors below the band gap energy. To quantify the contributions, we establish a computation scheme for the deflection signal. It essentially contains three steps: the computation of the sample's temperature field by finite element analysis, the computation of the refractive index field including the mirage effect, free carrier creation as well as pure field effects, and finally the semi-analytic computation of the probe beam path. The probe beam path yields the deflection signal and includes also the thermal expansion of the sample. To illustrate the significance of the different signal contributions we theoretically and experimentally study them in silicon at a pump laser wavelength of 1550 nm.

In silicon, the refractive index modulation in the bulk volume is mainly caused by a linear temperature induced refractive index change that leads to a computed parallel deflection of a few nanometers for typical experimental parameters. Noteworthy, there are several additional effects that are not directly linked to the absolute optical absorption but also contribute to the deflection. These are asymmetries in the refractive index field resulting from the chopper process and from spatial gradients of the attenuation coefficient along the probe beam axis which lead to an additional angular deflection of the probe beam. This causes a bulk signal whose magnitude depends on the distance of the readout point to the photo detector and may be misinterpreted as enhanced absorption.

Furthermore, for readout points close to the sample surface two additional effects change the refraction angle and thus the deflection of the probe beam: The thermally induced surface bulging of about 2 nm, experimentally verified by laser vibrometry and the bending of the probe beam because of the effects discussed above. The latter effect changes the refraction angle if the surface is located near the region where the probe beam is deflected. The surface related angular deflection can reach values of several micrometers and thus exceed the bulk deflection by far. This leads to an overestimation of the absorption if a sample surface is probed. With the computation scheme established in this work it is possible to account for all relevant spurious deflection effects and to extract the actual linear and nonlinear absorption data by measuring the PDS signal in dependence of the distance readout point – photodetector.

The measured silicon bulk absorption of about $0.02\ \text{cm}^{-1}$ corresponds well to literature values for the respective sample doping concentration.[24] The surface absorption is about ten times higher than the bulk value. The enhancement extracted from the distance dependent measurements is substantially less than reported in[16] which is particularly beneficial for applications in high-precision metrology where low noise is required.[5,6] The quadratic contribution to the silicon bulk deflection exceeds *5*% at a pump beam intensity of about $2.3 \times 10^5\ \text{W/cm}^2$. For comparison, the same threshold is experimentally determined in GaAs to be only $5 \times 10^4\ \text{W/cm}^2$ which is due to its larger two-photon absorption coefficient.[39]

In summary, this contribution provides guidelines how to evaluate key signal contributions in PDS and how to treat PDS measurement data to retrieve the optical absorption of semiconductors at photon energies below the bandgap with virtually arbitrary sample geometries. The presented method can provide additional information on the absorption processes at interfaces and in regions with spatial gradients in the absorption properties. We expect that this will provide valuable insights on the structural properties and absorption mechanisms of, e.g., [Si, SiO$_2$] multilayer systems[40,41] and semiconductor structured surfaces[19,20] and may thus pave the way for an optimization of their optical properties.

**ACKNOWLEDGEMENTS**

The authors gratefully acknowledge support by the Braunschweig International Graduate School of Metrology B-IGSM and the DFG research training group GrK1952/1 "Metrology for Complex Nanosystems".

# APPENDIX A: Temperature field and refractive index computation

In this appendix, we demonstrate details of the temperature and refractive index field computation. We provide explicit representations of terms for the finite element routine as well as values of relevant parameters.

We start with the intensity distribution of the pump beam that is directly linked to the heat source (see equation (3)). This intensity distribution is

$$I(\mathbf{x},t) = I_0 v(z) \left(\frac{w_0}{w(z-z_0)}\right)^2 \exp\left(-\frac{2(x^2+y^2)}{w(z-z_0)^2}\right) g(t) \quad (A1)$$

with the beam divergence

$$w(z) = w_0 \sqrt{1 + \left(\frac{z}{z_R}\right)^2} \quad (A2)$$

and the decay function for linear and quadratic absorption

$$v(z) = \frac{\alpha}{\exp(\alpha(z-z_{in}))(\alpha+\beta I_0) - \beta I_0}. \quad (A3)$$

Here, $I_0 v(z_0)$ is the pump beam intensity in the focus. The point $(0|0|z_{in})$ (see FIG. 3) indicates the entrance point of the pump beam into the sample. Furthermore, $w_0$ is the radius of the beam waist, $z_R$ is the Rayleigh range and $\beta$ is the 2-photon absorption coefficient. As there is an optical chopper modulating the pump beam intensity, a time modulation function $g(t)$ of the intensity must be considered.

The chopper process is illustrated in FIG. 13 (a). The pump beam radius at the passage through the chopper blades is $w_1$ and the opening width between the blades is $\Delta s$. The chopper frequency is $f_{Ch}$ and the pump duration is $N/f_{Ch}$. If the beam diameter $w_1$ is much smaller than the width of the blade $\Delta s$, the pump beam is almost completely blocked by the chopper wheel at certain moments. Rotating further to a distance $s$ between the chopper wheel edge and the beam centre, the following power $P_{trans}$ is transmitted:

$$P_{trans} = \int_{-\infty}^{\infty} \int_{-\infty}^{s} I_0 \times \exp\left(-2\frac{x^2+y^2}{w_1^2}\right) dx\, dy. \quad (A4)$$

Using equation (A11), the normalized transmitted power during the opening process follows. An analogical calculation leads to the normalized transmitted power during the chopper closing process:

$$\frac{P_{trans}}{P_0} = \frac{1}{2}\left[1 \pm \text{erf}\left(\frac{\sqrt{2}s}{w_1}\right)\right]. \quad (A5)$$

The time modulation $g(t)$ of the transmitted pump beam power follows by considering the chopper frequency:

$$g(t) = \sum_{j=0}^{N} \frac{1}{2}\left[1 + \text{erf}\left(\sqrt{2}\frac{\Delta s}{w_1}\left(2f_{Ch}t - \frac{4j+1}{2}\right)\right)\right] \times \text{rect}\left[2f_{Ch}t - \frac{4j+1}{2}\right] + \frac{1}{2}\left[1 - \text{erf}\left(\sqrt{2}\frac{\Delta s}{w_1}\left(2f_{Ch}t - \frac{4j+3}{2}\right)\right)\right] \times \text{rect}\left[2f_{Ch}t - \frac{4j+3}{2}\right]. \quad (A6)$$

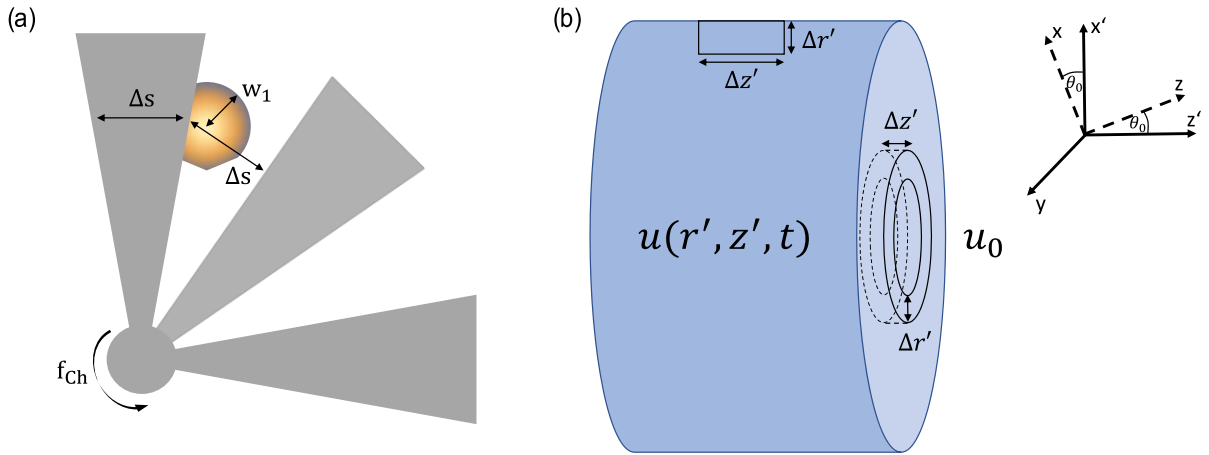

**FIG. 13.** (a) Shielding of the pump beam by an optical chopper. (b) Sketch of the surface volumes for the formulation of the boundary condition.

For the boundary conditions, the energy balance in an infinitesimal surface volume $V$ at the position $\boldsymbol{x}_s$ with area $A$ and depth $\Delta d$ ($\Delta d \to 0$) is drawn up (see FIG. 14 (b)):

$$\rho V c_\mathrm{m} \frac{\mathrm{d}u}{\mathrm{d}t} = -\lambda_\mathrm{Q} A \boldsymbol{\nabla} u \cdot \boldsymbol{N} + \alpha \times \iiint_V I(\boldsymbol{x}) dV \times g(t) - A\sigma\varepsilon(u^4 - u_0^4) - Ah(u - u_0)\,|_{\boldsymbol{x}=\boldsymbol{x}_s}. \quad (A7)$$

Here, $\boldsymbol{N}$ is the outward pointing normal vector of the sample surface, $\lambda_\mathrm{Q} = a\rho c_\mathrm{m}$ is the thermal conductivity, $\sigma$ is the Stefan-Boltzmann constant, $\varepsilon$ is the emissivity and $h$ is the heat transfer coefficient between the sample surface and the environment considering conduction and convection. On the left side of equation (A7), the net energy is expressed in terms of the resulting temperature change, while the right side contains the difference between incoming and outcoming energy. In the limit $\Delta d \to 0$ the left side of equation (A7) as well as the absorption term vanishes, and the boundary condition follows:

$$-\boldsymbol{\nabla} u(\boldsymbol{x},t)\cdot \boldsymbol{N} = \frac{\sigma\varepsilon}{\lambda_\mathrm{Q}}(u(\boldsymbol{x},t)^4 - u_0^4) + \frac{h}{\lambda_\mathrm{Q}}(u(\boldsymbol{x},t) - u_0)\,|_{\boldsymbol{x}=\boldsymbol{x}_s}. \quad (A8)$$

Some of the parameters occurring in the pump beam intensity are impractical to extract from the measurement setup. To provide a direct link to experimental parameters, the beam waist and the Rayleigh range are expressed in terms of the vacuum wavelength $\lambda_\mathrm{P}$, the pump beam radius at the converging lens $w_\mathrm{L}$, the distance between waist and incoming point $z_0 - z_\mathrm{in}$, the refractive index of the sample material in z-direction $n_z$ and the focal length $f_\mathrm{L}$ of the lens:

$$w_0 = \frac{\lambda_\mathrm{P}[n_z f_\mathrm{L} + (1-n_z)(z_0 - z_\mathrm{in})]}{\pi w_\mathrm{L}}, \quad (A9)$$

$$z_\mathrm{R} = \frac{\pi n_z w_0^2}{\lambda_\mathrm{P}}. \quad (A10)$$

The intensity $I_0$ is replaced by the easily measurable pump power $P_0$ by integration over the full beam cross section:

$$I_0 = \frac{2P_0}{\pi w_0^2}. \quad (A11)$$

The sample region is described as follows:

$$x'^2 + y^2 \leq R^2 \wedge z'\epsilon\left[-\frac{L}{2},\frac{L}{2}\right] \text{ with } x' = x\cos\theta_0 - z\sin\theta_0 \text{ and } z' = x\sin\theta_0 + z\cos\theta_0. \quad (A12)$$

The incoming position of the pump beam into the sample is given by

$$z_\mathrm{in} = -\frac{L}{2\cos\theta_0}. \quad (A13)$$

Since the pump beam radius is small compared to the sample radius, it is necessary to use a mesh which is finer close to the pump beam. Thus, the spatial dependent cell measure $\Delta R$ is described as a linear function of $\sqrt{x^2+y^2}$ with slope $m$ and minimal cell measure $\Delta R_0$:

$$\Delta R = m\sqrt{x^2+y^2} + \Delta R_0. \quad (A14)$$

Equations (A6) and (A9)-(A11) are inserted in equation (A1) leading to a very bulky representation of the pump beam intensity. For the sake of conciseness, we refrain from displaying it here. The input parameters for the computation of the temperature field are summarized in TABLE I. Additionally, TABLE II lists parameters for the refractive index field computation.

TABLE I. Input parameters for the temperature field computation.

| Quantity | Symbol | Value | Literature |
|---|---|---|---|
| thermal diffusivity | $a$ | $87 \times 10^{-6}$ m$^2$ s$^{-1}$ | 42 |
| mean attenuation coefficient | $\alpha$ | 2.86 m$^{-1}$ | measured |
| two-photon absorption coefficient | $\beta$ | $8 \times 10^{-12}$ m W$^{-1}$ | 39 |
| mass heat capacity | $c_\mathrm{m}$ | 700 J kg$^{-1}$ K$^{-1}$ | 43 |
| Stefan-Boltzmann constant | $\sigma$ | $5.670 \times 10^{-8}$ W m$^{-2}$ K$^{-4}$ | constant |
| emissivity | $\varepsilon$ | 0.70 | 44 |
| heat transfer coefficient | $h$ | 5 W m$^{-2}$ K$^{-1}$ | 45 |
| environment temperature | $u_0$ | 295.4 K | measured |
| volumetric mass density | $\rho$ | 2330 kg m$^{-3}$ | 46 |
| sample length | $L$ | 0.1 m | measured |
| sample radius | $R$ | 0.05 m | measured |
| pump power | $P_0$ | 5.0 W | measured |
| pump wavelength | $\lambda_\mathrm{P}$ | $1.55 \times 10^{-6}$ m | measured |
| refractive index sample | $n_z$ | 3.4757 | 46 |
| refractive index air | $n_0$ | 1.0003 | 47 |
| pump beam radius L | $w_\mathrm{L}$ | $2.30 \times 10^{-3}$ m | measured |
| chopper wheel opening width | $\Delta s$ | $8.00 \times 10^{-3}$ m | measured |
| chopper frequency | $f_\mathrm{Ch}$ | 108 s$^{-1}$ | measured |

TABLE II. Input parameters for the refractive index field computation.

| | | | |
|---|---|---|---|
| pump beam radius 1 | $w_1$ | $2.25 \times 10^{-3}$ m | measured |
| focal length | $f_L$ | 0.1 m | measured |
| cell measure slope | $m$ | 0.20 | defined |
| minimal cell measure | $\Delta R_0$ | $1.0 \times 10^{-5}$ m | defined |

| Quantity | Symbol | Value | Literature |
|---|---|---|---|
| probe wavelength | $\lambda$ | $1.31 \times 10^{-6}$ m | measured |
| electron lifetime | $\tau_C$ | $2.38 \times 10^{-4}$ s | 31 |
| effective electron mass | $m_e^*$ | $0.28\, m_e$ | 48 |
| effective hole mass | $m_h^*$ | $0.41\, m_e$ | 48 |
| Planck constant | $h$ | $6.626 \times 10^{-34}$ m$^2$ kg s$^{-1}$ | constant |
| vacuum speed of light | $c_0$ | $2.998 \times 10^8$ m s$^{-1}$ | constant |
| vacuum permittivity | $\varepsilon_0$ | $8.854 \times 10^{-12}$ F m$^{-1}$ | constant |
| elementary charge | $e$ | $1.602 \times 10^{-19}$ C | constant |
| electron mass | $m_e$ | $9.109 \times 10^{-31}$ kg | constant |
| oscillator resonance frequency | $\omega_0$ | $2\pi \times 10^{15}$ s$^{-1}$ | 29 |
| average oscillator displacement | $X$ | $10^{-9}$ m | 29 |
| thermo-optic coefficient | $dn/du$ | $1.94 \times 10^{-4}$ K$^{-1}$ | 27 |

**APPENDIX B: Ray tracing in the bulk**

This section outlines the main steps for the semi-analytic ray tracing yielding an expression for the deflection $d_{bulk}$. The deflection is the maximum beam displacement of the probe beam leaving the sample. The schematic probe beam path is illustrated in FIG. 14.

The slope angle of the probe beam path $z = F(x,t)$ is the sum $\varphi(\vec{x},t) + \varphi_N(\vec{x},t)$ (compare FIG. 15), where $\varphi_N(x,t)$ is the angle between the x-axis and the normal vector on the local equipotential of the refractive index field. The differential equation for $F(x,t)$ is:

$$F(x,t): \frac{dz}{dx}(t) = \tan(\varphi(\vec{x},t) + \varphi_N(\vec{x},t)). \quad (B1)$$

Note that $F(x,t)$ is independent of the y-coordinate, as pump and probe beam propagate in the y=0 plane. The angle $\varphi_N(x,t)$ is computed with the refractive index field:

$$\tan(\varphi_N(x,t)) = \frac{\partial_z n(x,t)}{\partial_x n(x,t)}. \quad (B2)$$

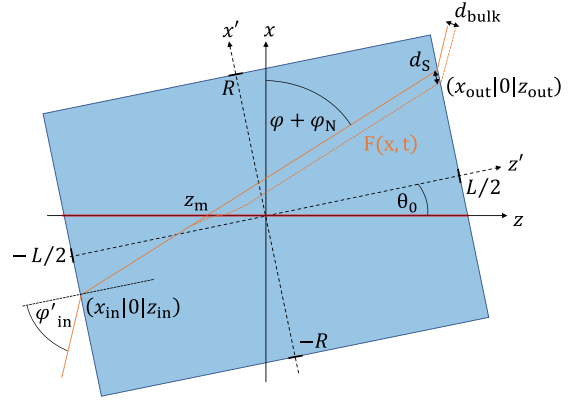

**FIG. 14.** Sketch of the probe beam path through the sample. The probe beam enters the sample of length $L$ from the left in the entrance point $(x_{in}|0|z_{in})$, crosses the pump beam axis in $(0|0|z_m)$ and leaves the sample in $(x_{out}|0|z_{out})$. Due to the periodic deflection process the outcoming periodically changes its position. The maximum distance between these outcoming points is $d_S$, while the maximum probe beam displacement behind the back surface is $d_{bulk}$. The latter value is detected by the quadrant diode (compare FIG. 1).

Using equations (B2) and (9), equation (B1) is integrated. The result is given in equation (10). The integration constant is calculated with the entrance point of the probe beam

$$x'_{in} = \left[\left(\frac{L}{2\cos\theta_0} + z_m\right)\left(1 - \frac{n_0^2}{n_z^2}\sin^2\varphi'_{in}\right)^{-1/2} \sin\left[\theta_0 + \arcsin\left(\frac{n_0}{n_z}\sin\varphi'_{in}\right)\right] - \frac{L}{2}\left(\frac{1}{\cos^2\theta_0} - 1\right)^{1/2}\right]e_{x'} - (L/2)e_{z''}, \quad (B3)$$

where the conversion to beam coordinates is defined by equation (A12). The incidence angle of the probe beam relative to the z'-axis is $\varphi'_{in}$. Using equation (B3), the integral in equation (10) and thus the probe beam path is defined without a free integration constant, as $x'_{in} \in F(x,t)$. Probing far away from the surface, thermal expansion effects at the entrance point can be neglected. In this case the numerical aperture is

$$NA(t) = n(x_{in},t) \cdot \sin\left(\frac{\pi}{2} - \theta_0 - \arcsin\left(\frac{n_0 \sin\varphi'_{in}}{n(x_{in},t)}\right) - \varphi_N(x_{in},t)\right). \quad (B4)$$

The deflection $d_{\text{bulk}}(t)$ is finally computed with the beam displacement $d_S(t)$ between the intersections $x_{\text{out}}(t)$ of $F(x,t)$ and the back surface of the sample. The intersection point is computed as:

$$x_{\text{out}}(t): \frac{L}{2} - x_{\text{out}}(t) \cdot \sin\theta_0 \equiv F(x_{\text{out}}(t),t) \cdot \cos\theta_0. \quad (B5)$$

The deflection reads as follows:

$$d_{\text{bulk}}(t) = d_S(t)\cos\varphi'_{\text{in}} \quad (B6)$$

with

$$d_S(t) = \left[(x_{\text{out}}(t) - x_{\text{out}}(0))^2 + \left(F(x_{\text{out}}(t),t) - F(x_{\text{out}}(0),0)\right)^2\right]^{1/2}. \quad (B7)$$

**APPENDIX C: Surface signal computation**

In the PDS setup, two effects lead to a modulation of the refraction angle at the surface: The thermal expansion and contraction of the sample and the probe beam bending due to the index gradient field (see FIG. 5 (b)). The influence of both effects is illustrated in FIG. 15 for the front surface. At the back surface analogue effects occur. The surface slope angular change $\chi$ due to thermal expansion is defined by the slope of the surface shape change $z'(x')$. The transformed shape $z(x)$ is used to replace $dz'/dx'$ by the easier derivative $d(\Delta z_{\text{max}})/dx$:

$$z(x) = \Delta z_{\text{max}}(x) - x \cdot \tan\theta_0, \quad (C1)$$

with

$$\Delta z_{\text{max}}(x) = \Delta\hat{l} \cdot \exp\left[-\frac{x^2}{\sigma_z^2}\right]. \quad (C2)$$

The corresponding fit parameters are given below FIG 6. The slope angle of the undisturbed surface $z(x)$ is obviously $-\theta_0$. Together with the disturbance due to the thermal expansion, the total slope angle is $\chi-\theta_0$. The corresponding slope reads as

$$\frac{dz}{dx} = \frac{d}{dx}(\Delta z_{\text{max}}) - \tan\theta_0 = \tan(\chi-\theta_0). \quad (C3)$$

The small surface slope angle ($\chi \ll 1$) allows for the approximation $\chi = d(\Delta z_{\text{max}})/dx$. The total probe beam angle change $\gamma_{\text{bulk}}$ in the sample centre due to index gradient bending is given by the slope of the probe beam path $d_{\text{bulk},0}(p)$ illustrated in FIG. 5 (b) at the incoming point by $\tan\gamma_{\text{bulk}} = d(d_{\text{bulk},0})/dp$. Here, $p$ is the probe axis position that denotes the distance from the crossing point between the two beams along the probe beam propagation direction. The probe beam path $d_{\text{bulk},0}(x)$ is computed out of $d_{\text{bulk},0}(p)$ by the transformation $x = p \times \sin(2\theta_0)$ and fitted as

$$d_{\text{bulk},0}(x) = 2.04 \text{ nm} - \frac{4.08}{1+\exp(-8.10x)}. \quad \text{(C4)}$$

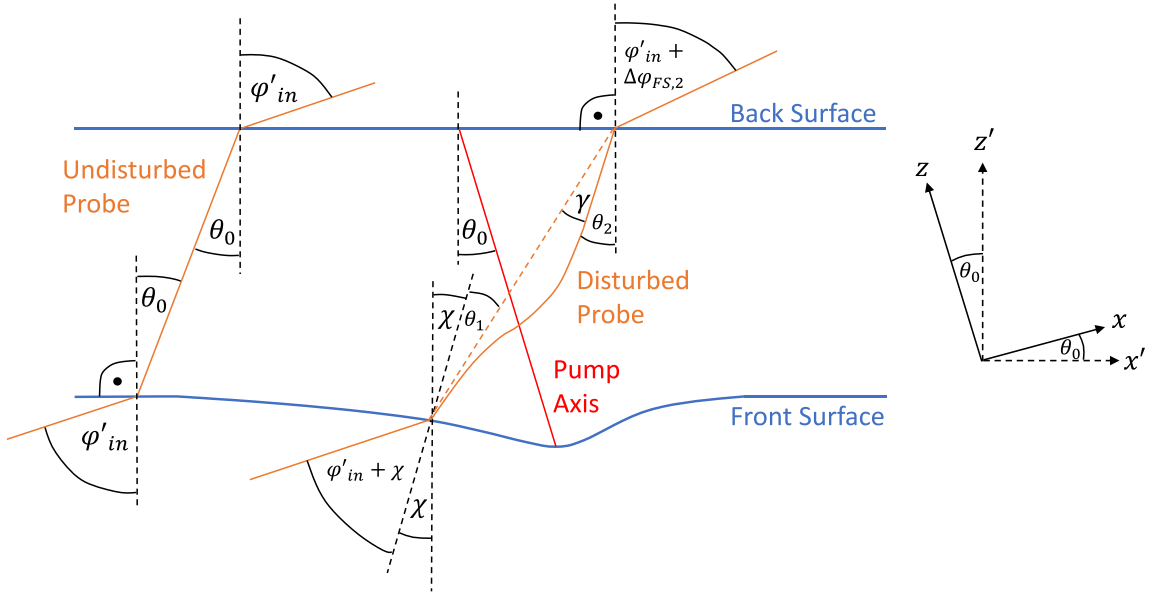

**FIG. 15.** Schematic probe beam paths through the sample for an undisturbed (chopper completely closed) and disturbed case (chopper completely open). The surface slope angle due to thermal expansion is $\chi$ and the total probe beam angular change in the sample due to index gradient bending is $\gamma$.

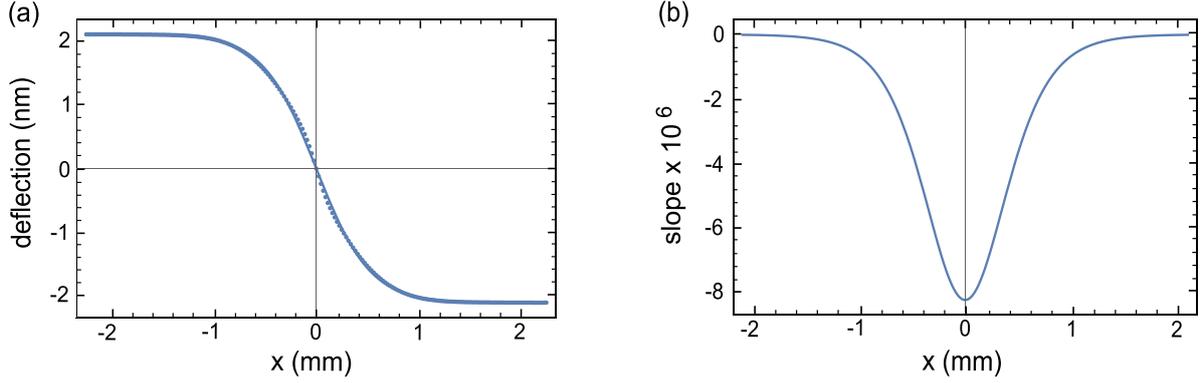

**FIG. 16.** (a) The fitted probe beam path $d_{\text{bulk},0}(x)$ and (b) its derivative.

The fit function and its derivative are plotted in FIG. 16. Using this result and again small angle approximation, the angle change is computed like $\gamma_{\text{bulk}} = \sin(2\theta_0) \cdot d(d_{\text{bulk},0})/dx$. Taking into account the attenuation of the pump beam during its way through the sample, the angle change at the front surface is increased and at the back surface is reduced by a factor of $\exp[\alpha L/(2\cos\theta_0)]$ compared to the sample centre. Since both $\chi$ and $\gamma$ are known as a function of $x$, the next step is the computation of $\Delta\varphi_{\text{FS},1} = \theta_2 - \theta_0$ with $\theta_2 = \theta_1 - \chi + \gamma$ (angle change at the front surface after the index gradient bending) and $\Delta\varphi_{\text{FS},2}$ (back surface). The refraction law at the front surface reads as

$$n_0 \sin(\varphi'_{\text{in}} + \chi) = n_z \sin\theta_0. \quad \text{(C5)}$$

Using the analogous relation for the undisturbed case $n_0 \sin\varphi'_{\text{in}} = n_z \sin\theta_0$, the relation $\theta_1 = \theta_0 + \Delta\varphi_{\text{FS},1} + \chi - \gamma$, small angle approximations ($\chi \ll 1$, $\Delta\varphi_{\text{FS},1} \ll 1$, $\gamma \ll 1$) and addition theorems, the resulting angle change in the sample reads as

$$\Delta\varphi_{\text{FS},1} = \chi\left(\frac{n_0 \cos\varphi'_{\text{in}}}{n_z \cos\theta_0} - 1\right) + \gamma. \quad \text{(C6)}$$

The refraction law at the back surface reads as

$$n_z \sin\theta_2 = n_0 \sin(\varphi'_{\text{in}} + \Delta\varphi_{\text{FS},2}). \quad \text{(C7)}$$

Using the same relations as above and $\theta_2 = \Delta\varphi_{\text{FS},1} + \theta_0$, the angle change behind the back surface follows:

$$\Delta\varphi_{\text{FS},2} = \frac{n_z \cos\theta_0}{n_0 \cos\varphi'_{\text{in}}} \Delta\varphi_{\text{FS},1}. \quad \text{(C8)}$$

With the present experiment configuration $\varphi'_{\text{in}} = \pi/2 - \theta_0$, equation (C6) is equivalent with (13b) and (C8) with (13c). The corresponding angular deflection in (13a) is a sum of the angular deflection in the sample and behind the back surface. The readout position $D_m$ relates to $x$ by:

$$x = D_m \times \tan(\theta_0) \quad \text{(C9)}$$

The computation of the back surface deflection given in (14a) is conducted analogically.

**APPENDIX D: Extraction of the attenuation coefficient out of the PDS signal**

We discuss the extraction of the attenuation coefficient from the corrected deflection signal $g(z)$ for two important cases: First, we assume a neglectable quadratic contribution to the deflection signal which is valid for low intensities and semiconductors with small two-photon absorption coefficients (compare section IV C). In this case, $g(z)$ obeys the following relation:

$$g(z) \propto \frac{dP(z)}{dz} = -\alpha(z)P(z). \quad \text{(D1)}$$

Here, $P(z)$ denotes the total pump power at a given position $z$. The solution of this Bernoulli-type differential equation is

$$P(z) = P(z_{\text{in}})\exp\left[-\int_{z_{\text{in}}}^{z} \alpha(w)dw\right] \quad \text{(D2)}$$

with the entrance position $z_\text{in}$. Using equation (D2) in (D1) and calculating the derivative of $g(z)$, we get

$$\frac{g'(z)}{g(z)} = \frac{\alpha'(z)}{\alpha(z)} - \alpha(z). \quad \text{(D3)}$$

The solution of equation (D3) is given by equation (15), using the total transmittance $T$ to calculate the integration constant. For high pump beam intensities and semiconductors with high two-photon absorption coefficients, the power dependence of the PDS signal (compare section IV C) should be evaluated to check the influence of a quadratic contribution. If this contribution is not neglectable, then $g(z)$ obeys:

$$g(z) \propto \frac{\mathrm{d}P(z)}{\mathrm{d}z} = -\alpha(z)P(z) - \beta(z)P^2(z). \quad \text{(D4)}$$

The solution of this Bernoulli-type differential equation then is

$$P(z) = \frac{\exp\left[-\int_{z_\text{in}}^{z}\alpha(w)\mathrm{d}w\right]}{1/P(z_\text{in}) + \int_{z_\text{in}}^{z}\left(\beta(v)\exp\left[-\int_{z_\text{in}}^{z}\alpha(w)\mathrm{d}w\right]\mathrm{d}v\right)}. \quad \text{(D5)}$$

Again, we calculate the derivative of $g(z)$ to get

$$\frac{g'(z)}{g(z)} = \frac{\alpha'(z) - \alpha^2(z) + \left(\beta'(z) - 3\alpha(z)\beta(z)\right)P(z) - 2\beta^2(z)P^2(z)}{\alpha(z) + \beta(z)P(z)}. \quad \text{(D6)}$$

Using equation (D6) and (D4), we eliminate the unknown power function $P(z)$ and get the differential equation for $\alpha(z)$ and $\beta(z)$:

$$\left(\alpha(z)\frac{g'(z)}{g(z)} + \alpha^2(z) + 2\beta(z)g(z) - \alpha'(z)\right)\left(\alpha(z)\beta'(z) + 2\beta^2(z)d(z) - \beta(z)\alpha'(z)\right) = g(z)\left(\beta'(z) - \alpha(z)\beta(z) - \beta(z)\frac{g'(z)}{g(z)}\right)^2. \quad \text{(D7)}$$

This equation must be solved numerically requiring at least two PDS signals $g(z)$ at two different powers to compute $\alpha(z)$ and $\beta(z)$.